\documentclass{elsart}
\usepackage{graphics}
\begin{document}
\begin{frontmatter}
\hyphenation{Coul-omb ei-gen-val-ue ei-gen-func-tion Ha-mil-to-ni-an
  trans-ver-sal mo-men-tum re-nor-ma-li-zed mas-ses sym-me-tri-za-tion
  dis-cre-ti-za-tion dia-go-na-li-za-tion in-ter-val pro-ba-bi-li-ty
  ha-dro-nic he-li-ci-ty Yu-ka-wa con-si-de-ra-tions spec-tra
  spec-trum cor-res-pond-ing-ly}
\title{B-physics phenomenology with emphasis on the light-cone}
\author{Chueng-Ryong Ji and Ho-Meoyng Choi}
\address{Department of Physics, North Carolina State University,
Box 8202, Raleigh, NC 27695-8202}
\date{\today}
\begin{abstract}
Theoretical overviews on the B-physics are presented 
with an emphasis on the light-cone degrees of freedom. 
Our new treatment of the embedded states seems to give an
encouraging result.
\end{abstract}
\end{frontmatter}
\section{Introduction}
\label{introduction}
With the wealth of new and upgraded experimental facilities for the
B-physics, the precision test of standard model is ever more promising.
The new facilities of BaBar at SLAC and Belle at KEK generate asymmetric
$e^+ e^-$ collisions to trace the B-decays more accurately. The symmetric
$e^+ e^-$ collision facility of CLEO-III at Cornell has also upgraded
its luminosity by a factor of ten better. Furthermore, the hadon-hadron 
collision facilities such as LHCB at CERN and BTeV at Fermilab as well 
as the lepton-hadron collision facility such as HERA-B at DESY emerge 
as the powerful tools to investigate a lot of detailed B-decays~\cite{Sh}.
Certainly, an important motivation to study B-physics is to make
a precision test of standard model especially associated with the 
unitarity of CKM mixing matrix. One of the burning questions in physics
is whether the complex phase is really the only source of the 
CP-violation or not.
To make such a precision test, the accurate analyses of
exclusive semileptonic B-decays as well as rare B-decays are strongly
demanded. 
As we will discuss in this talk, an effective use of light-cone (LC)
degrees of freedom in those analyses seem crucial to make the 
calculations more accurate. This also makes the model-building 
more scrutinized. Our talk is presented with the following outlines.
The theoretical overview of B-physics is given in the next Section,
Section 2. Especially, we will discuss the processes determining
each CKM-matrix element and the profile of unitary triangle.
We'll also try to make a very brief survey of theoretical development
in the last twenty years history of B-physics. Because of enormous works
that people have done in B-physics, this survey would be in no way complete
but very brief and limited. In any case, the purpose of Section 2 
will be to motivate why the exclusive semileptonic B-decays are very 
important to constrain the CKM mixing matrix. In Section 3, then
exclusive semileptonic decays are discussed. Some general remarks on
the weak form factors will be made and the role of LC degrees of
freedom in the exclusive semileptonic decays will be discussed.
Especially, the difficulties associated with the time-like processes
such as the exclusive semileptonic decays in the LC formulation.
We will first identify the embedded states necessary to restore the
covariance of the amplitudes and then present a way of handling the
embedded states. Some preliminary numerical results are also presented
in the exclusive semileptonic decays for $K\to\pi\ell\nu$ with our new way
of treating the embedded states. Conclusions follow in Section 4.    
\section{The Theoretical Overview of B-Physics}
\label{overview}
In the standard model, the only interaction relevant
to the CKM mixing matrix $V_{CKM}$ is the weak charged-current interaction
given by the Lagrangian ${\mathcal L}_{\rm int}=-(g/\sqrt{2})(W^+_\mu J^\mu 
+ W^-_\mu J^{\mu \dagger})$, where
\begin{equation}
J_\mu = \left(\bar{u}, \bar{c}, \bar{t} \right) 
\gamma_\mu V_{CKM} \left(\begin{array}{c} d\\s\\b \end{array} \right),\;\;
{\rm and}\;\; 
V_{CKM} = \left(\begin{array}{ccc}
V_{ud} & V_{us} & V_{ub} \\
V_{cd} & V_{cs} & V_{cb} \\
V_{td} & V_{ts} & V_{tb} \end{array} \right).
\label{eq:ckm}
\end{equation}
In the standard model, $V_{CKM}$ is unitary ,i.e. 
$V_{CKM}^\dagger V_{CKM} = 1$, and all the matrix elements of $V_{CKM}$
can be written in terms of three real angles $\theta_{12}, \theta_{23},
\theta_{13}$ and one real phase-angle $\delta$ as explicitly shown 
in the particle data group~\cite{PDG}. 
Here, $\theta_{12}$ is the usual Cabbibo
mixing angle. Wolfenstein~\cite{Wolf} realized the pattern of
order of magnitude in each element and parametrized $V_{CKM}$ 
in the orders of $\lambda = \sin\theta_{12} \approx 0.22$. Up to 
the order of $\lambda^3$, $V_{CKM}$ is given by
\begin{equation}
V_{CKM}\approx \left(\begin{array}{ccc}
1-\lambda^2/2 & \lambda & A\lambda^3 (\rho - i\eta) \\
-\lambda & 1-\lambda^2/2 & A\lambda^2 \\
A\lambda^3 (1-\rho-i\eta) & -A\lambda^2 & 1
\end{array} \right),
\label{eq:wolfenstein}
\end{equation}
where four real Wolfenstein's parameters are given by
($A,\lambda,\rho,\eta$). 

The unitarity condition leads to the definition of various
unitarity triangles:
\begin{eqnarray}
V^\dagger_{CKM}V_{CKM} = \left(\begin{array}{ccc}
(dd) & (ds)^* & (db)^* \\
(ds) & (ss)   & (sb)^* \\
(db) & (sb)   & (bb) \end{array}\right),
V_{CKM}V^\dagger_{CKM} = \left(\begin{array}{ccc}
(uu)   & (uc)   & (ut) \\
(uc)^* & (cc)   & (ct) \\
(ut)^* & (ct)^* & (tt) \end{array} \right)
\label{eq:vvdagger}
\end{eqnarray}
where each off-diagonal element $(ij)$ with the unequal quark-flavors 
$i$ and $j$, i.e. $i\not =j$, is
given by the addition of three complex numbers due to the multiplication
of two 3$\times$3 complex matrices, e.g. 
$(db) = V_{ud}V_{ub}^* + V_{cd}V_{cb}^* + V_{td}V_{tb}^*$, 
while the diagonal elements are given by the sum of 
three real numbers. Since the off-diagonal element $(ij)$ ($i\not = 
j$) is zero (i.e. the sum of three complex numbers is zero), it can be
given as a triangle in the complex plane. Thus, each off-diagonal element
$(ij)$ corresponds to a unitarity triangle and there are six independent
ones. Only four out of eighteen angles in the six triangles are 
independent and the area of all triangles is identical measure
of CP-violation, i.e., ${\rm Area}(\Delta) 
= (1/2)\sin\theta_{12} \sin\theta_{23} \sin\theta_{13} 
\sin\delta \approx(1/2)A \lambda^6 \eta$.
Also, the magnitude of each mixing matrix element $|V_{ij}|$ is independent
of parametrization.

Now, let's focus on the current magnitude~\cite{PDG} of each mixing matrix 
element and the corresponding experimental processes 
used for the determination.
The magnitude of $V_{ud}$ has been determined by the superallowed 
$0^+\rightarrow0^+$ nuclear $\beta$-decay, the nucleon 
$\beta$-decay ($n\rightarrow p+e+\bar{\nu_e}$) and the
pion $\beta$-decay ($\pi^+\rightarrow\pi^0 + e^+ +\nu_e$). The current 
average value is given by $|V_{ud}|=0.9735\pm0.0008$.
The magnitude of $V_{us}$ and $V_{cd}$ is almost equal to the sine of
the Cabbibo angle, i.e. $|V_{us}|=0.2196\pm0.0023$ and 
$|V_{cd}|=0.224\pm0.016$, respectively. $|V_{us}|$ is determined by
the semileptonic kaon decay $K_{l3}$ ($K\rightarrow \pi l \nu$) and
the hyperon semileptonic decays such as $\Lambda\to p e\bar{\nu_e},
\Sigma^- \rightarrow n e \bar{\nu_e}, \Xi \rightarrow \Lambda e \bar{\nu_e}$,
and $\Xi \rightarrow \Sigma^0 e \bar{\nu_e}$, while $|V_{cd}|$ is determined
by the semileptonic $D$-meson decays of $D^0 \rightarrow \pi^- e^+ \nu_e$,
$D^+ \rightarrow \pi^0 e^+ \nu_e$ as well as the leptonic decays 
$D^+ \rightarrow\mu^+ \nu_\mu, D^- \rightarrow\mu^-\bar{\nu_\mu}$.
The magnitude of $|V_{cs}| = 1.04\pm0.16$ is also determined by the $D$-meson
semileptonic decays $D^0\rightarrow K^- e^+ \nu_e, D^+\rightarrow 
\bar{K^0}e^+\nu_e$. The analysis of heavy-to-heavy semileptonic $B$-decays 
such as $B^+ \rightarrow \bar{D^0} l^+ \nu_l$ and $B^+ \rightarrow 
\bar{D^0}^* l^+\nu_e$ has been constrained by the heavy quark effective 
theory (HQET)~\cite{IW,HQET} and the small value of $|V_{cb}|$ was 
determined as $|V_{cb}|=0.0402\pm0.0019$. The HQET has also played the 
role of constraining the model building. Even smaller value of 
$|V_{ub}|=(0.090\pm0.025)|V_{cb}|$ was determined
by the heavy-to-light semileptonic $B$-decays of $B\rightarrow \pi l \nu_l,
\rho l \nu_l, \omega l \nu_l$ as well as the leptonic decays of 
$B^+\rightarrow \mu^+ \nu_\mu, \tau^+ \nu_\tau$. 
In this way, the top two rows of $V_{CKM}$ matrix were determined 
mostly by the direct measurements of experimental processes.

However, the direct measurements are not feasible for the bottom row 
elements of $V_{CKM}$ matrix involving $t$-quark. Since the $t$-quark mass
is so heavy ($m_t = 174.3\pm5.1$ GeV) and the lifetime of $t$-quark is much
shorter than the strong interaction time scale, the $t$-quark doesn't 
have any time to form a bound-state meson but quickly decays
into $b$-quark and $W^+$. Nevertheless, the rough magnitudes of 
$|V_{tb}|\approx 1, |V_{ts}|\approx 0.04, |V_{td}|\approx 0.005\sim0.013$
are consistent with the semileptonic decays of $t$-quark 
$t\rightarrow(b,s,d)l^+ \nu_l$ measured at CDF and D$\emptyset$ in the 
Fermilab where the $t$-quark evidence was confirmed. Especially, 
the constraint given by
\begin{equation}
\frac{|V_{tb}|^2}{|V_{td}|^2+|V_{ts}|^2+|V_{tb}|^2} = 0.99\pm0.29
\label{eq:t-ratio}
\end{equation}
is well satisfied by these magnitudes. 
While the direct measurements are not feasible for $V_{tb},V_{ts},V_{td}$,
the virtual transition through ``Penguin" process will give more 
informations. 
The first evidence of ``Penguin" process was seen at CLEO II~\cite{CLEO2}, 
where the branching ratio of radiative $B$-decay was determined as
BR($B\rightarrow K^*(892) \gamma$) = ($4.2\pm0.8\pm0.6)\times10^{-5}$.

\begin{figure}
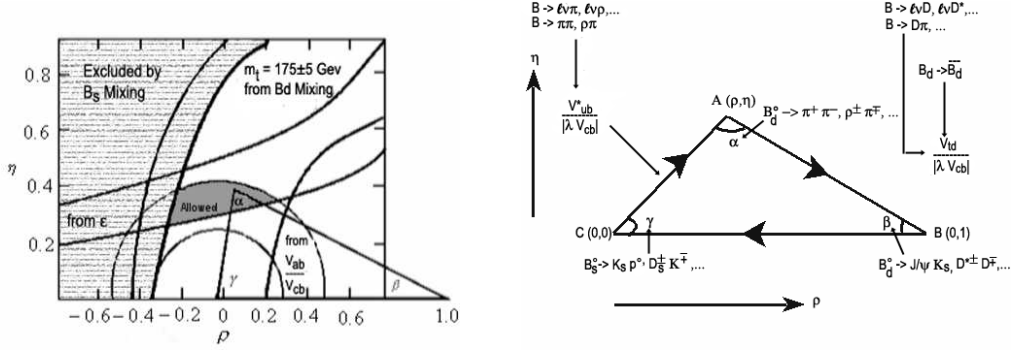

\begin{center}
\resizebox{1.0\textwidth}{!}{
 \includegraphics{trianglefinalrev.eps}
\hspace{2cm}
 \includegraphics{trianglegreek.eps}
}\caption{\label{fig:1} 
(a) Allowed region for the apex ($\rho,\eta$) of the unitary triangle.
(b) Processes to determine the profile of unitary triangle.
}\end{center}
\end{figure}

The most interesting and promising unitarity triangle to be measured
is (db);
\begin{equation}
V_{ud} V_{ub}^* + V_{cd} V_{cb}^* + V_{td} V_{tb}^* = 0,
\label{eq:triangle}
\end{equation}
where $V_{ud}\approx 1, V_{cd}\approx -\lambda, V_{tb}^*\approx 1$.
Dividing the l.h.s. of Eq.~(\ref{eq:triangle}), one gets
\begin{equation}
\frac{V_{ub}^*}{|\lambda V_{cb}|} -1 + \frac{V_{td}}{|\lambda V_{cb}|} =0,
\label{eq:triangle-prime}
\end{equation}
where each term can be written in terms of Wolfenstein's parameters 
$\rho$ and $\eta$ in the complex plane of ($\rho, \eta$). The first term in 
Eq.(\ref{eq:triangle-prime}) corresponds to the vector from (0,0) to 
($\rho,\eta$) and the last term does the vector from ($\rho,\eta$) to (0,1). 
The second term $-1$ corresponds to the vector from (0,1) to (0,0). 
The main interest in determining the profile 
of the unitarity triangle is then to find the exact location of the apex 
($\rho,\eta$). The allowed region of the apex is given by the overlapping
region in Fig.~\ref{fig:1}(a). The length between the two apices (0,0) and 
($\rho,\eta$)
is mainly determined by the measurement of ratio $V_{ub}/V_{cb}$.
The allowed region for this length is given by the rainbow around
(0,0) in Fig.~\ref{fig:1}(a). 
Also, the length of the adjacent side connecting 
($\rho,\eta$) to (1,0) is essentially determined by the measurement of 
$B_d$ mixing that can provide the value of $V_{td}$. Since the value of 
$V_{cb}$ is very close to that of $V_{ts}$, one can further constrain 
this length between ($\rho,\eta$) to (1,0) by measuring the $B_s$ mixing. 
The rainbow around (1,0) shown in Fig.~\ref{fig:1}(a) corresponds to the 
allowed region for this length. 
The overlap of the two rainbows is further constrained
by the measurement of CP-violation parameter $\epsilon$ in $K^0$-$\bar{K^0}$ 
mixing. Since $\epsilon$ is a direct measure of the complex phase in 
$V_{CKM}$, the constraint from $K^0$-$\bar{K^0}$ mixing is drawn as a 
horizontal band in Fig.~\ref{fig:1}(a) depending mostly on the value of $\eta$.
The final overlap yields then the allowed region for the apex ($\rho,\eta$)
of the triangle as shown in Fig.~\ref{fig:1}(a). The three angles of the 
triangle  denoted by $\alpha, \beta,\gamma$ are given by
\begin{eqnarray}
\alpha = {\rm arg}\left(-\frac{V_{td} V_{tb}^*}{V_{ud} V_{ub}^*}\right),
\beta = {\rm arg}\left(-\frac{V_{cd} V_{cb}^*}{V_{td} V_{tb}^*}\right),
\gamma = {\rm arg}\left(-\frac{V_{ud} V_{ub}^*}{V_{cd} V_{cb}^*}\right).
\label{eq:angles}
\end{eqnarray}
It is very interesting to note that these angles can also be determined
from the CP-asymmetry measurement~\cite{Nir}. 
The CP-asymmetry is given by
\begin{eqnarray}
A_{f_{CP}}(t) &=& \frac{\Gamma(B^0_{\rm phys}(t)\to f_{CP}) 
- \Gamma(\bar{B^0}_{\rm phys}(t)\to f_{CP})}{\Gamma(B^0_{\rm phys}(t)
\to f_{CP}) + \Gamma(\bar{B^0}_{\rm phys}(t)\to f_{CP})},\\
\nonumber
&{\approx}& 
- {\rm Im}\left(\frac{q}{p}\cdot\frac{\bar{\mathcal M}}{\mathcal M}\right)
\sin(\Delta M t), 
\label{eq:asymmetry}
\end{eqnarray}
where $|B_L\rangle = p |B^0\rangle + q |\bar{B^0}\rangle$,
$|B_H\rangle = p |B^0\rangle - q |\bar{B^0}\rangle$,
${\mathcal M} = \langle f_{CP}|{\mathcal H}| B^0\rangle$, and
$\bar{{\mathcal M}} = \langle f_{CP}|{\mathcal H}|\bar{B^0}\rangle$.
From the experimental measurements of $A_{f_{CP}}(t)$, it may be
possible to determine the values of angles $\alpha, \beta, \gamma$.
For example, the value of $\beta$ may be determined by the measurement
of $B_d^0\to J/\psi K_s$.
More examples of process to determine each angle are shown in 
Fig.~\ref{fig:1}(b). Figure~\ref{fig:1}(b) also shows other processes
to fix the profile of the unitary triangle.

The theoretical efforts in the $B$ physics have also been
very extensive in the last twenty years and we cannot summarize all the 
developments in this talk. Perhaps, here we are just content
with a very brief survey. In early 80's, the value of $V_{cb}$ was
investigated significantly with the QCD improved spectator model developed by
Altarelli et al.~\cite{Alt}. In the middle of 80's, a model wavefunction
due to Wirbel, Stech and Bauer called WSB model~\cite{WSB} was 
introduced for the heavy quark system. Then, in the late 80's, the heavy
quark symmetry was extensively studied by many authors~\cite{IW,HQET}. 
Perhaps, the most quoted work has been done by Isgur and Wise~\cite{IW} 
introducing the Isgur-Wise function now frequently referred
in the heavy quark effective theory 
(HQET). The basic idea of heavy quark symmetry is to realize the 
hidden symmetry of QCD that can be revealed only in the limit of
infinitely heavy quark masses. The analogous observation in the opposite
extreme of zero quark mass limit is the chiral symmetry. Whether the quark
masses are heavy or light may be categorized by the QCD scale 
$\Lambda_{QCD}$. The current quark masses of $u,d,s$ are much smaller
than $\Lambda_{QCD}$, while the masses of $c,b,t$ quarks are much larger
than $\Lambda_{QCD}$, {\em i.e.} $m_u , m_d, m_s << \Lambda_{QCD} << 
m_c,m_b,m_t$. Thus, in the limit that $m_u, m_d$ and $m_s$ go to zero,
the QCD reveals the chiral symmetry $SU(3)_L\times SU(3)_R$ and the symmetry
is spontaneously broken to $SU(3)_V$ due to the non-trivial QCD vacuum.
Similarly, in the limit that $m_c,m_b$ and $m_t$ go to infinity,
the heavy quark symmetry $SU(6)$(or $SU(2N_f)$) is revealed in QCD. 

In the early 90's, the HQET was applied extensively in the heavy quark
systems. At the same time, the constituent quark model(CQM) was built by
Isgur, Scora, Grinstein and Wise and named as 
ISGW model~\cite{ISGW1}. This model
emphasizes the importance of resonance contribution near the small
invariant mass of final states in the inclusive semileptonic decays.
In the case of semileptonic $B$ decay such as $B\rightarrow e \nu c d$,
the invariant mass square $P_X^2=m_X^2=(p_B-p_e-p_\nu)^2$ of the final 
state, continuum $c$ and $d$ quarks, is bound by 
$(m_c+m_d)^2 < P^2_X < (m_c+m_d)^2+ (m_d/m_b)(m_b-m_c)$,
where $p_e, p_\nu$ and $p_B$ are the four-momenta of electron, neutrino
and the $B$ meson, respectively. 
ISGW criticized the early QCD improved spectator
model by Altarelli {\em et. al.}~\cite{Alt} because the region of
$P_X^2 \approx (m_c+m_d)^2$ is dominated by the resonances rather than
the continuum. In the middle of 90's, ISGW model was extended to a 
relativistic version and called ISGW2 model~\cite{ISGW2}. 
Throughout 90's, the lattice QCD~\cite{Lat1} was also extensively 
used to analyze the heavy quark systems.
In 90's, the perturbative QCD(PQCD)~\cite{pQCD} approach was used mainly 
to analyze the hadronic decays of heavy mesons. 
Then, in the later part of 90's, dispersion relation was used for 
the analysis of timelike region using the inputs from CQM~\cite{Mel}, 
lattice data~\cite{Bec}, HQET and PQCD~\cite{Boyd}. The light-cone quark
model(LCQM) was developed around this time. Perhaps, we may categorize
the LCQM into four different versions. First, the ISGW2 model was extended
to the LC formalism~\cite{LF1}. However, the same input parameters as the 
ISGW2 model were used in this LCQM. 
Second, the LC version of HQET was developed~\cite{LF2}.
In this development, the WSB model developed in the middle of 80's 
was ruled out because the WSB model doesn't satisfy the constraint from
the heavy quark symmetry. Third, the LC model wavefunction was also
introduced~\cite{LF3}. However, modelling the LC wavefunction didn't give 
any information about the hadron spectra. Finally, we have implemented the
variational principle to the QCD motivated effective LC hamiltonian
to enable the analysis of meson spectra as well as many 
wavefunction-related observables such as the form factors, decay constants 
and electroweak decay rates, etc.~\cite{CJ1,CJ2}. 
In late 90's, also other approaches
such as the light-cone QCD sum-rule~\cite{sum}, the Dyson-Schwinger 
equation~\cite{DS} and the Bethe-Salpeter equation~\cite{BS} were used 
to analyze the heavy quark systems. 
More recently, the quark-meson model~\cite{Gato} utilizing both HQET and
chrial perturbation ($\chi_{PT}$) theory was used to analyze the 
heavy-to-light meson decays such as $B\rightarrow (\pi,\rho,a_1)\nu l$.
In the twenty years of $B$-physics history, we may note that
one of the focal point in most analyses has been the accurate prediction 
of exclusive semileptonic decays such as $B\rightarrow (D,\pi) l\nu$
and $D\rightarrow (K,\pi) l\nu$. In the next section, we now present
some more details of exclusive semileptonic decays.
\section{Exclusive Semileptonic Decays}
\label{overview2}
The current matrix element of the semileptonic pseudoscalar to 
pseudoscalar (PS$\to$PS) meson decays involve the two form factors:
\begin{eqnarray}
\langle p_2|V^\mu|p_1\rangle &=& 
f_+(q^2)(p_1+p_2)^\mu + f_-(q^2)(p_1-p_2)^\mu \\
\nonumber
&=& f_+(q^2)\biggl[(p_1+p_2)^\mu 
- \frac{M_1^2 - m_2^2}{q^2} q^\mu\biggr]
+ f_0(q^2)\frac{M_1^2-M_2^2}{q^2}q^\mu\\
\nonumber
&=& \sqrt{M_1 M_2}[h_+(q^2)(v_1+v_2)^\mu + h_-(q^2)(v_1-v_2)^\mu],\\
\label{eq:ps to ps}
\end{eqnarray}
where
\begin{equation}
f_0(q^2)=f_+(q^2)+\frac{q^2}{M_1^2-M_2^2}f_-(q^2).
\end{equation}
In the limit of $M_{1,2}\rightarrow\infty$, the only one form factor
remains, {\em i.e.} $h_+(q^2)=\xi(v_1 \cdot v_2)$ and $h_-(q^2)=0$,
where $\xi(v_1 \cdot v_2)$ is known as the Isgur-Wise function.
Also, due to the Ademello-Gatto theorem~\cite{AG}, $f_+(0)\approx 1$. 
Note here that $f_+(q^2)$ corresponds to the charge form factor in the 
$SU(3)$ limit.
The similar decomposition of the matrix element of the semileptonic
pseudoscalar (PS) to vector (V) meson decays can be made with the four
transition form factors. Again in the heavy quark mass limit, only one form
factor $\xi(v_1\cdot v_2)$ is needed. Also, the Luke theorem~\cite{Luke} 
applies in the zero recoil limit. However, because of the limited space, 
we won't discuss the details of PS$\to$V semileptonic decays in this 
presentation but here focus only on the PS$\to$PS semleptonic decays. 

The analysis of exclusive processes can be made efficiently in the LC 
formalism with the rational energy-momentum relation. In the LCQM calculations 
presented in Ref.~\cite{LF1}, the $q^+\neq0$ frame has been used to calculate
the weak decays in the timelike region $m^2_\ell\leq q^2\leq (M_1-M_2)^2$,
with $M_{1(2)}$ and $m_\ell$ being the initial (final) meson mass and the
lepton ($\ell$) mass, respectively. However, when the $q^+\neq0$ frame is
used, the inclusion of the nonvalence contributions arising from 
quark-antiquark pair creation (``Z-graph") is inevitable and this inclusion
may be very important for the heavy-to-light and light-to-light decays.
Nevertheless, the previous analyses~\cite{LF1} in the $q^+\neq0$ frame 
considered only valence contributions neglecting the nonvalence 
contributions. 

\begin{figure}
\begin{center}
\resizebox{0.8\textwidth}{!}{
 \includegraphics{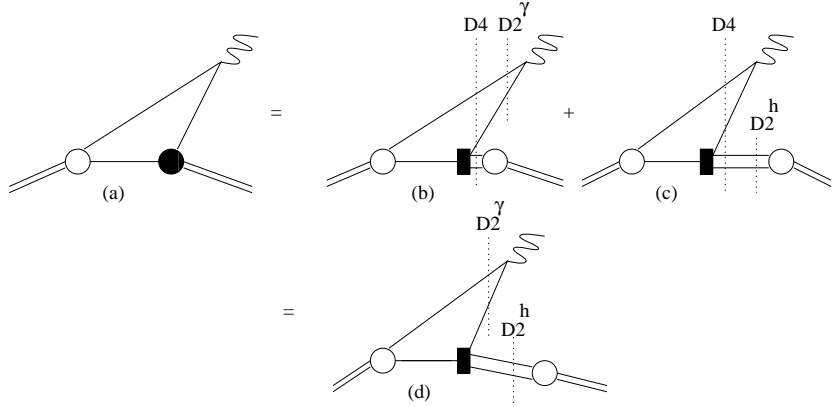}
}\caption{\label{fig:Embed}
Effective calculation of the embedded state (black blob) in terms
of the usual LC wave fucntion (white blob).
}\end{center}
\end{figure}
In this work, we treat the nonvalence state using the Schwinger-Dyson
equation to connect the embedded-state shown as the black blob
in Fig.~\ref{fig:Embed} to the ordinary LC wave function (white blob in 
Fig.~\ref{fig:Embed}). To make the program successful, we need some relevant
operator connecting one-body to three-body sector shown as the black box 
in Fig.~\ref{fig:Embed}. 
The relevant operator is in general dependent on the involved momenta.
Our main observation is that we can remove the four-body energy denomenator
$D_4$ using the identity $1/D_4D^\gamma_2 + 1/D_4D^h_2=1/D^\gamma_2D^h_2$
of the energy denominators and obtain the identical amplitude in terms of
ordinary LC wave functions of photon and hadron (white blob). 
For the small momentum transfer, perhaps the relevant operator 
may not have too much dependence on the involved
momenta and one may approximate it as a constant operator.
In contact interaction case, we verified that our prescription of a 
constant operator in Fig.~\ref{fig:Embed}(d) is an exact solution of 
Fig.~\ref{fig:Embed}(a).
In our previous analysis~\cite{CJ2} for the exclusive
PS$\to$PS semileptonic decays, the form factor $f_{+}(q^2)$ obtained
from $j^+$ in $q^{+}=0$ frame is not only immune to the zero-mode
contribution but also in good agreement with the experimental data
as well as other theoretical results. In order to obtain the form factor
$f_-$ in $q^+=0$ frame, one has to use another component (i.e. $j_{\perp}$
or $j_-$) of the current in addition to the $j^+$.
However, as noted in Ref.~\cite{Ja1}, those $j_{\perp}$ and $j_{-}$
are not immune to the zero-mode contributions, which are not easy to be
identified in LCQM. Thus, we use $q^+\neq0$ frame to determine the constant 
operator by equating the slope of $f_+$ at $q^2=0$ in $q^+\neq0$ frame to 
that in $q^+=0$ frame. Then, we apply the same operator to the calculation 
of $f_-$. We present here some preliminery results for $K_{\ell3}$ using the
approximation of constant operator to illustrate our method.
\begin{table*}
\begin{center}
\caption{\label{kpi} 
Preliminary results for the parameters of $K^{0}_{\ell3}$ decay
form factors.}
\begin{tabular}{|c|c|c|c|c|}\hline
 &\multicolumn{2}{c|}{$q^+\neq0$ frame}& $q^+=0$ frame& \\
\cline{2-4}
 &Effective(val + nv) &valence & valence & Experiment~\cite{PDG}\\
\hline
$f_{+}(0)$ & 0.962 & 0.962 & 0.962 & \\
\hline
$\lambda_{+}$& 0.026 & 0.083 & 0.026 & $0.0288\pm0.0015[K^{0}_{e3}]$\\
\hline
$\lambda_{0}$& 0.025 & $-0.017$& 0.001
& $0.025\pm0.006[K^{0}_{\mu3}]$\\
\hline
$\xi_{A}$& $-0.013$ & $-1.10$& $-0.29$
& $-0.11\pm0.09[K^{0}_{\mu3}]$\\
\hline
\end{tabular}
\end{center}
\end{table*}
In Table~\ref{kpi}, we summarize the experimental observables for the 
$K_{\ell3}$ decays, where $\lambda_i=M^{2}_{\pi}f'_{i}(0)/f_{i}(0)(i=+,0)$ 
and $\xi_{A}=f_-(0)/f_+(0)$. We use our linear potential parameters
given by Refs.~\cite{CJ1,CJ2} in this analysis.
As one can see in Table~\ref{kpi}, our results for the slope
$\lambda_0$ of $f_0$ at $q^2=0$ and $\xi_{A}$=$f_-(0)/f_+(0)$ 
are now much improved and comparable with the data. 
Especially, our result of $\lambda_0$= 0.025 
obtained from our effective calculation is in excellent 
agreement with the data, $\lambda^{\rm Exp.}_0$=0.025$\pm$0.006.
More theoretical details on our effective method as well as heavy-to-heavy
and heavy-to-light semileptonic processes will be presented in 
the future communication.
\section{Conclusion}
With the wealth of new or upgraded experimental facilities, precision
test of standard model is ever more promising. We presented a theoretical
overview of B-physics and noted that the exclusive semileptonic B-decays 
and rare B-decays are very important to determine 
$V_{bc}$, $V_{bu}$, $V_{td}$, $V_{ts}$, and $V_{tb}$ stringently. 
As we discussed, an effective use of LC degrees of freedom 
seems crucial to make predictions consistent with many other exclusive 
processes. Our initial attempt to accomodate the contributions from 
embedded states seems to give encouraging results. 

We would like to thank Prof. Chris Pauli for organizing a stimulating
meeting and his outstanding hospitality at the Max Plank Institute.
We also thank Chirag Lakhani for drawing a figure. This work was supported
by a grant from the U.S. DOE under contracts DE-FG02-96ER40947.

\end{document}